\documentclass[10pt]{article}
\usepackage{graphicx,amssymb,amsmath,subfigure,fullpage}
\usepackage{ucs}
\usepackage[utf8x]{inputenc}

\newtheorem{definition}{Definition}

\begin{document}

\begin{center}
{\bf {\Large Yet Another Proof of the Aperiodicity of Robinson Tiles}}\\
~\\
{\large Thomas~Fernique}
\end{center}

\begin{figure}[hbtp]
\centering
\includegraphics[width=0.8\textwidth]{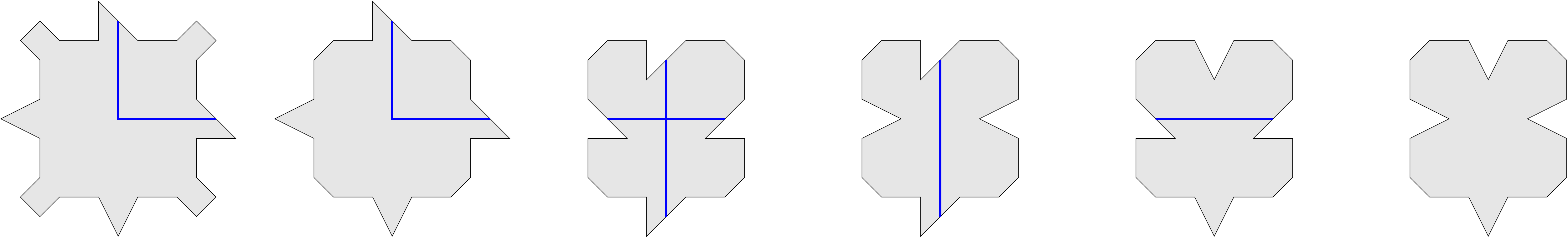}
\caption{\small The six Robinson tiles (they can be rotated or reflected).
The leftmost tile is said to be \emph{bumpy}, the other ones \emph{dented}.
Triangular notching of tile edges are called \emph{arrow}.
An arrow is oriented \emph{inwards} or \emph{outwards}, depending on whether the notching is a dent or a bump, and can have two \emph{types}, depending on whether it meets a blue line or not.
The two leftmost tiles are called \emph{corner}, the other ones \emph{arm}.
An arm is oriented towards its unique outwards arrow (here, four South-arm) and a corner towards the concavity of its blue corner (here, two NorthEast-corners).
}
\label{fig:robinson_tileset}
\end{figure}

An \emph{aperiodic tile set} is a finite set of tiles which can tile the plane (\emph{i.e.} cover it without overlap), but not in a periodic way (\emph{i.e.}, no tiling is invariant by a translation).
To prove the undecidability of the \emph{the Domino problem}, Robert Berger constructed in 1966 the first aperiodic tile set \cite{berger}.
His proof was simplified in 1971 by Rafael Robinson, who introduced a six-tile set and proved its aperiodicity.
Fig.~\ref{fig:robinson_tileset} depicts these celebrated six tiles and introduces some terminology.\\

\begin{figure}
\centering
\includegraphics[width=0.9\textwidth]{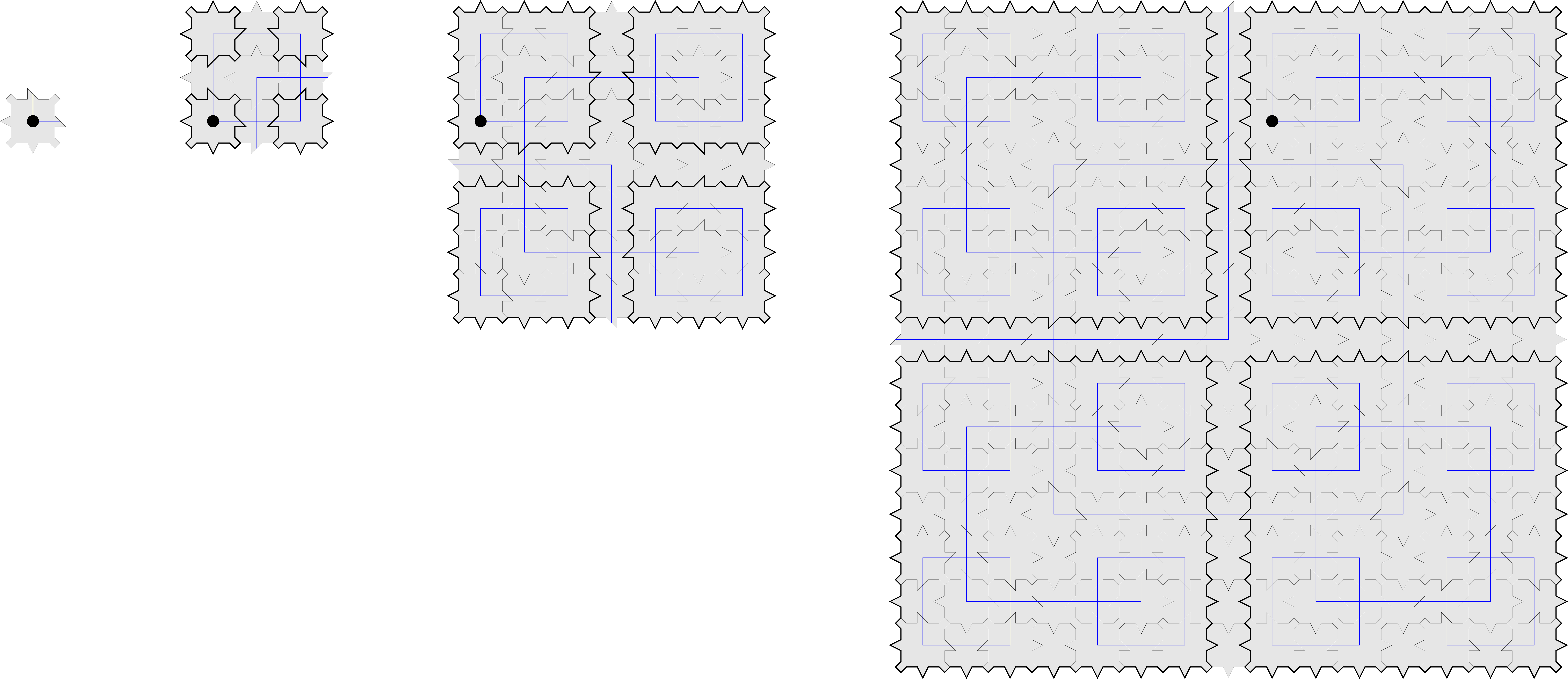}
\caption{\small 
Order-increasing bumpy corners whose orientation turns clockwise yield a sequence of spiral-growing tilings
which converges towards a tiling of the whole plane
(left-to-right: order $1$--$4$ bumpy corners oriented $NE$, $SE$, $SW$ and $NW$; the black dot marks the origin and order $n-1$ bumpy corners appearing in order $n$ ones are emphasized).
}
\label{fig:generalized_bumpy_corners}
\end{figure}

\noindent To prove the aperiodicity of this tile set, we inductively define \emph{generalized bumpy corners} as follows:

\begin{definition}
The usual bumpy corner is said to be of order $1$.
Then, for $n\geq 2$, the order $n$ bumpy corner is the square tiling made of four order $n-1$ bumpy corners oriented towards a central dented corner, with arms filling the gaps\footnote{One checks that this can be done in a unique way.} (Fig.~\ref{fig:generalized_bumpy_corners}).
The bumpy corners of order $n\geq 2$ are oriented as their central dented corner.
\end{definition}

On the one hand, consider a sequence of order-increasing bumpy corners whose orientation turns clockwise (Fig.~\ref{fig:generalized_bumpy_corners}).
This is an increasing sequence of finite tilings which cover arbitrarily big balls around the first bumpy corner .
By putting at a given position of the plane the tile which eventually appears at this position in this sequence\footnote{Since once a tile appears in a tiling of this sequence, it appears at the same position in the following tilings.}, we get a tiling of the whole plane.
Thus, Robinson tilings do exist.\\

\begin{figure}[hbtp]
\centering
\subfigure[Consider a bumpy NE-corner of order $n$.]{\includegraphics[width=0.2\textwidth]{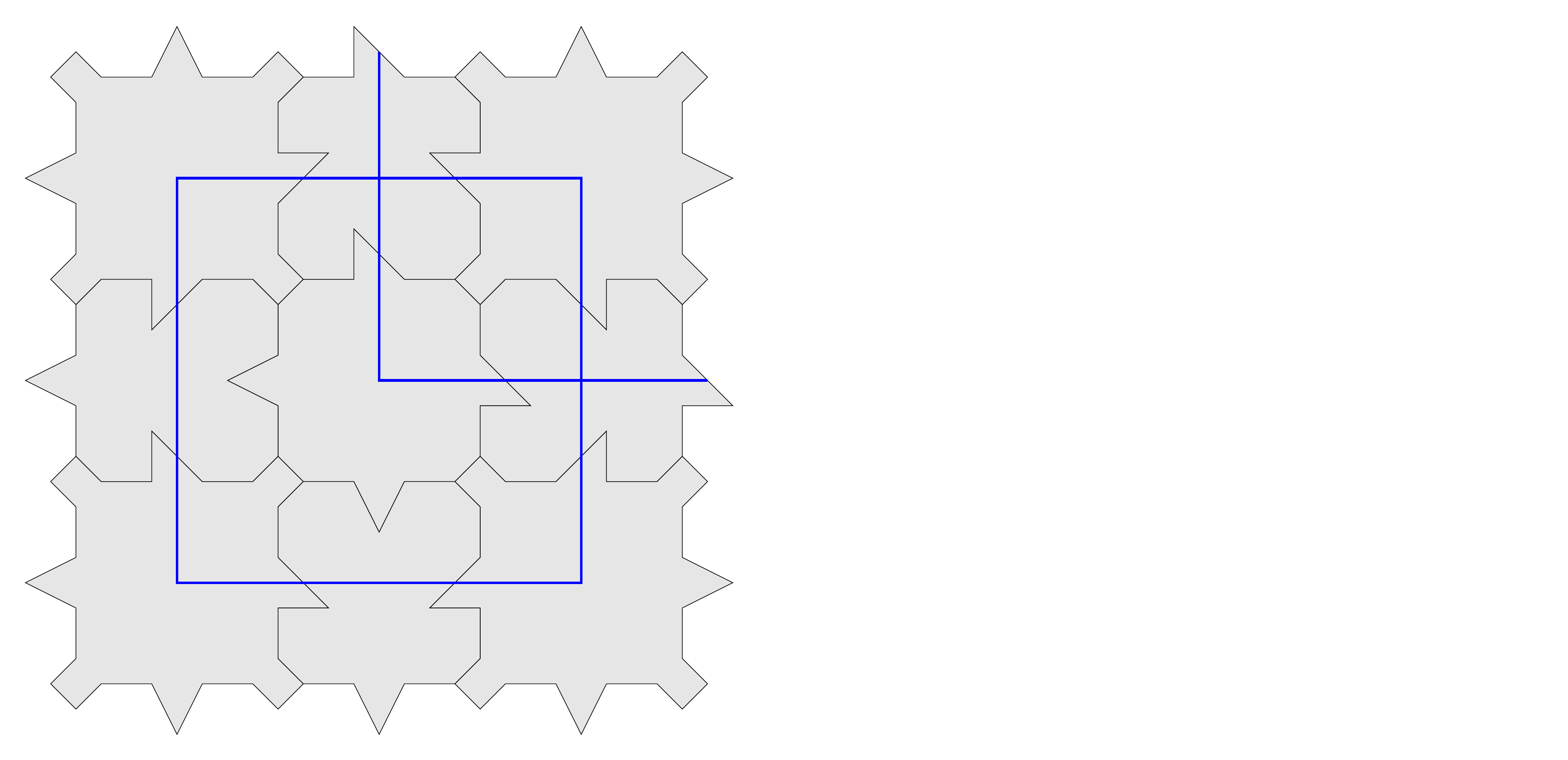}}\hfill
\subfigure[The tiles along its east side can only be arms.]{\includegraphics[width=0.2\textwidth]{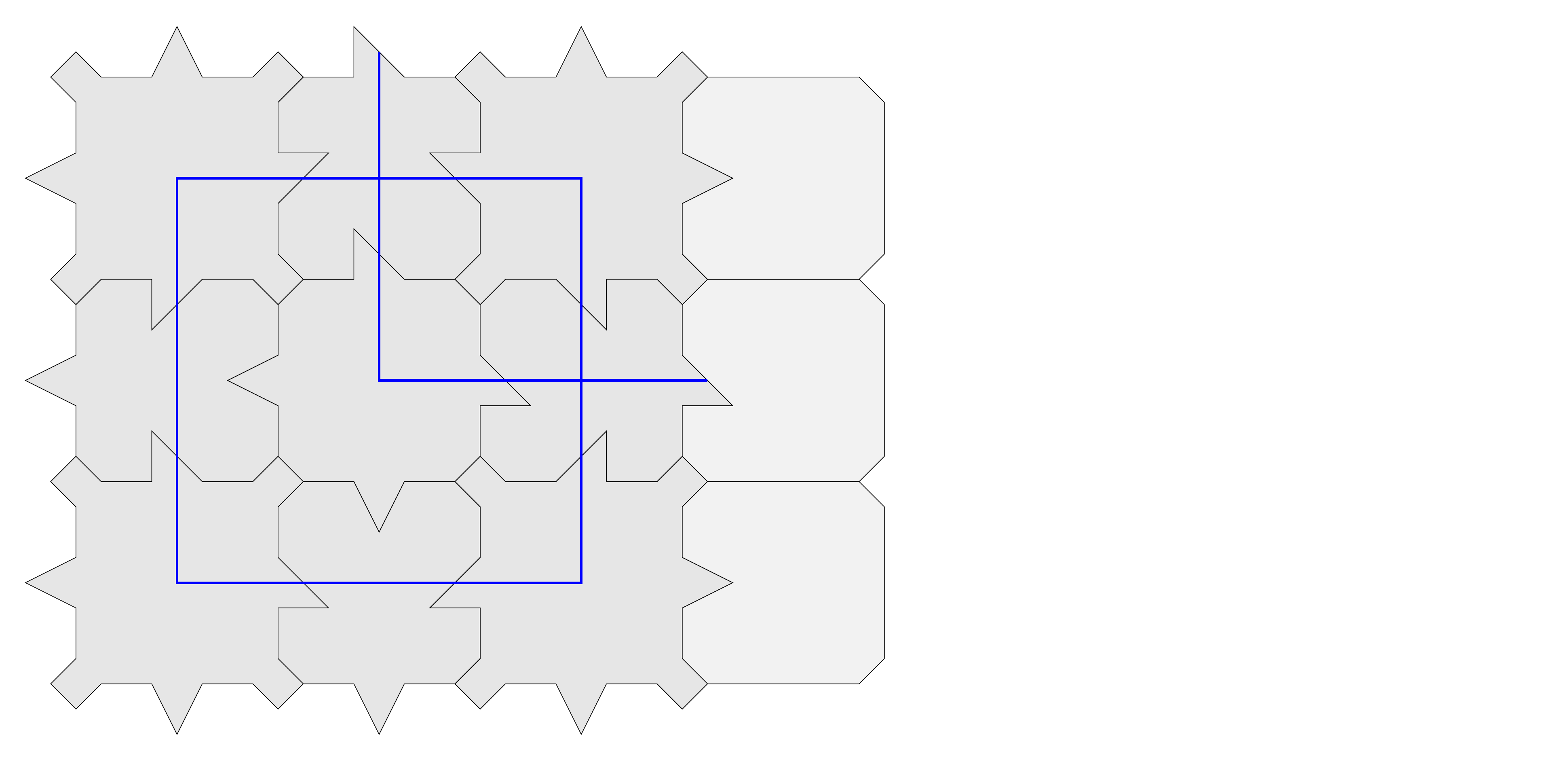}}\hfill
\subfigure[The middle one is S or E: its N-arrow is inwards.]{\includegraphics[width=0.2\textwidth]{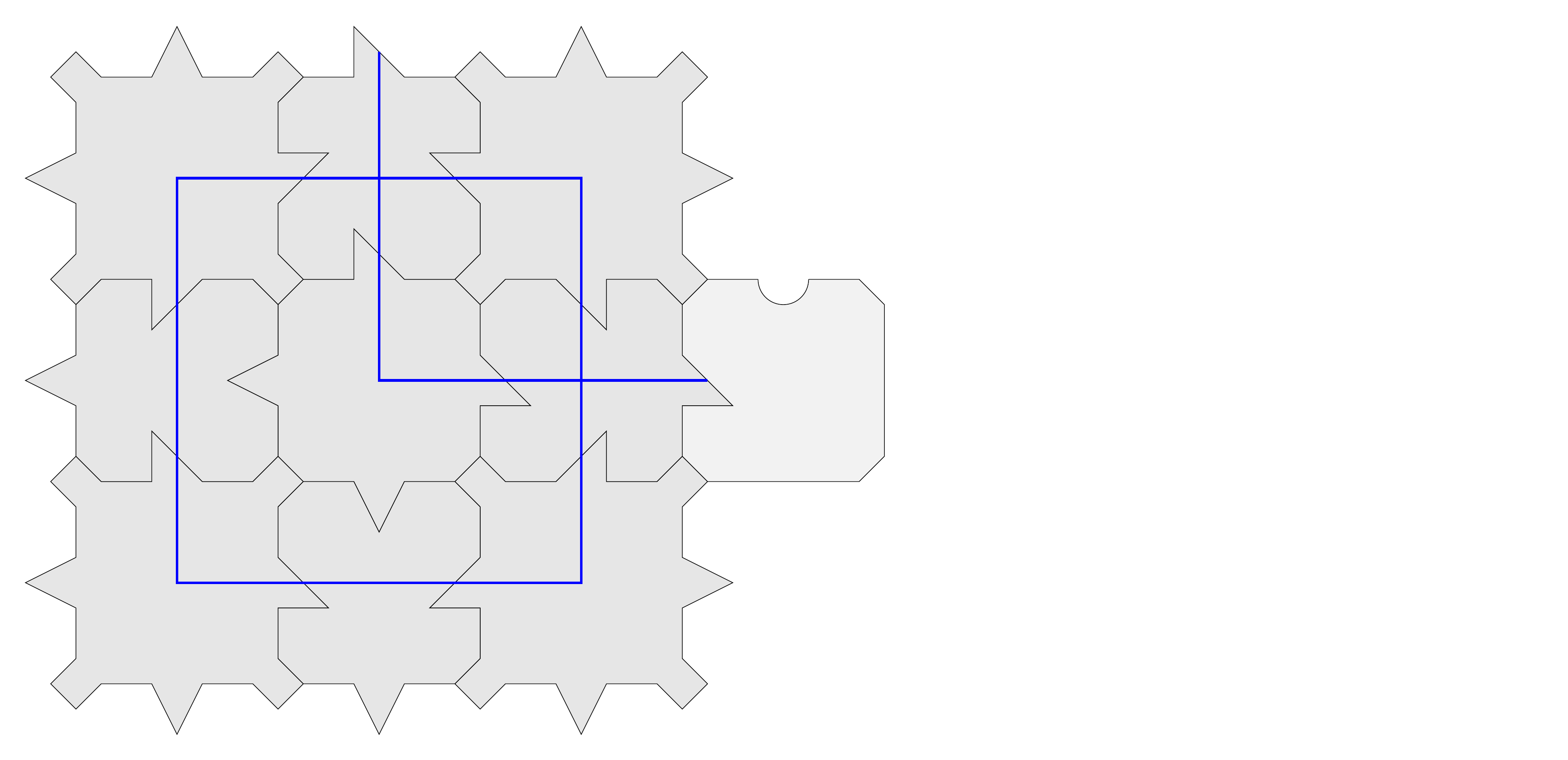}}\hfill
\subfigure[This forces northern arms to be S-arms.]{\includegraphics[width=0.2\textwidth]{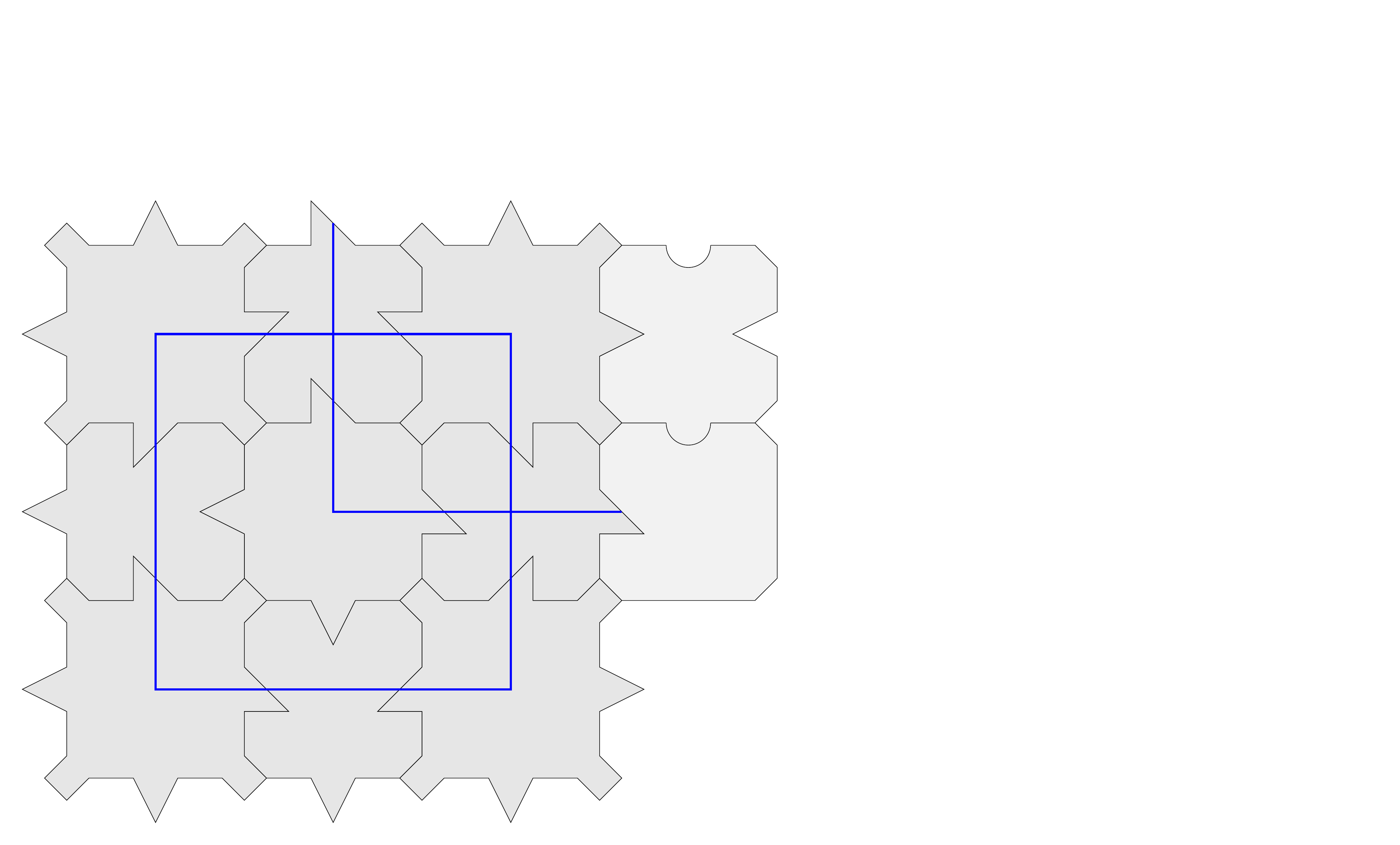}}\\
\vspace{-1.2cm}
\subfigure[Symmetrically north.]{\includegraphics[width=0.2\textwidth]{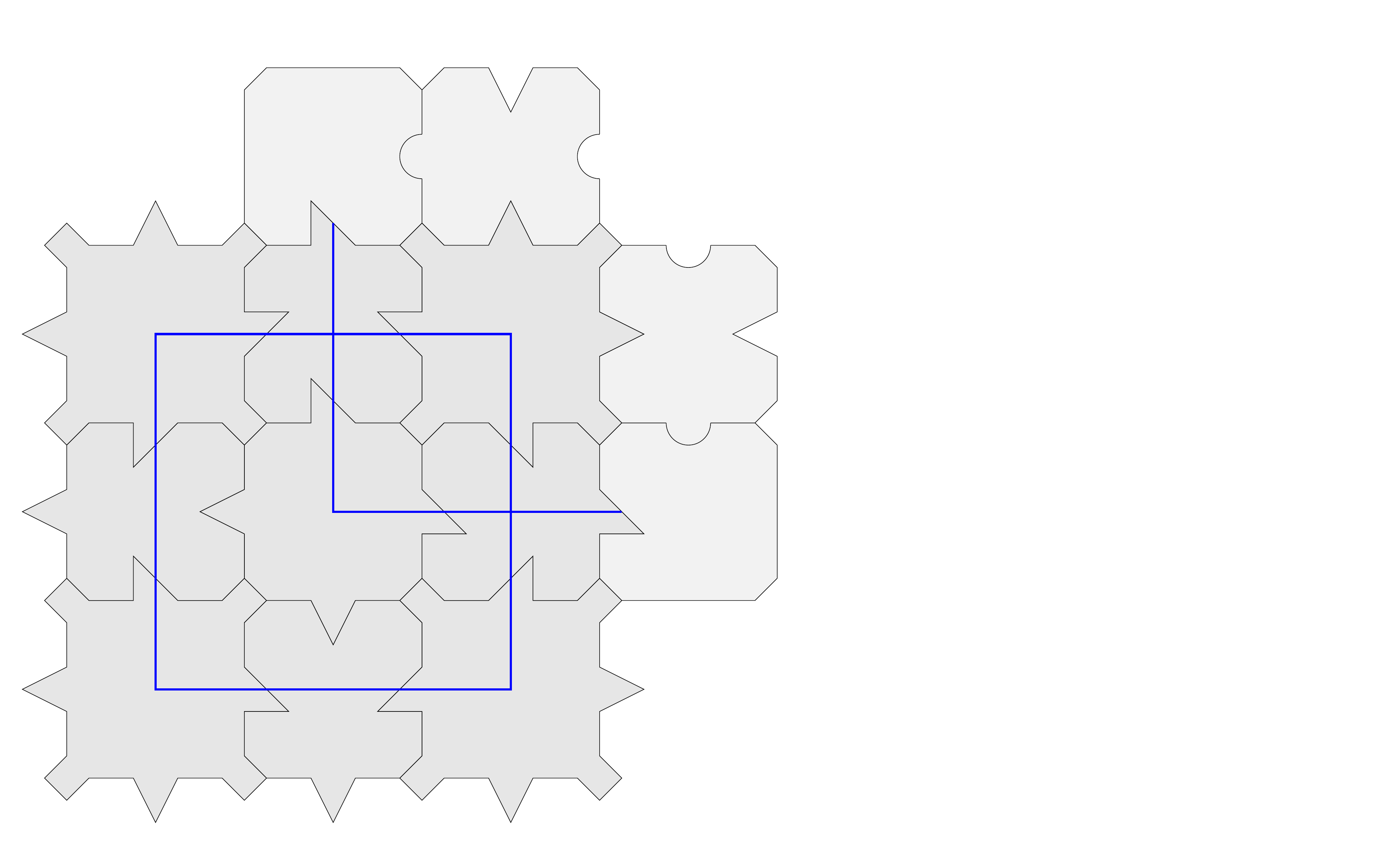}}\hfill
\subfigure[This forces a corner,]{\includegraphics[width=0.2\textwidth]{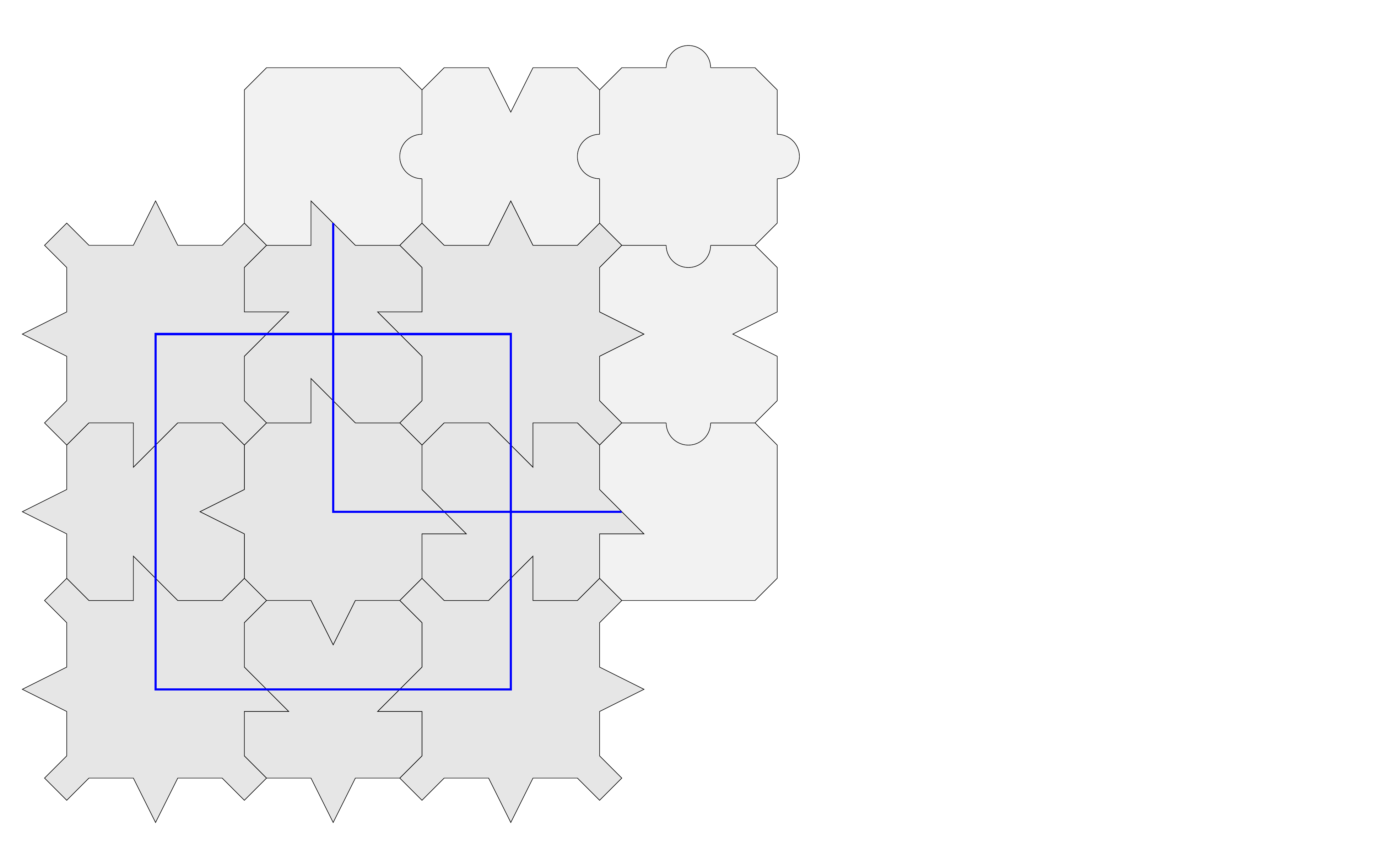}}\hfill
\subfigure[two arms,]{\includegraphics[width=0.2\textwidth]{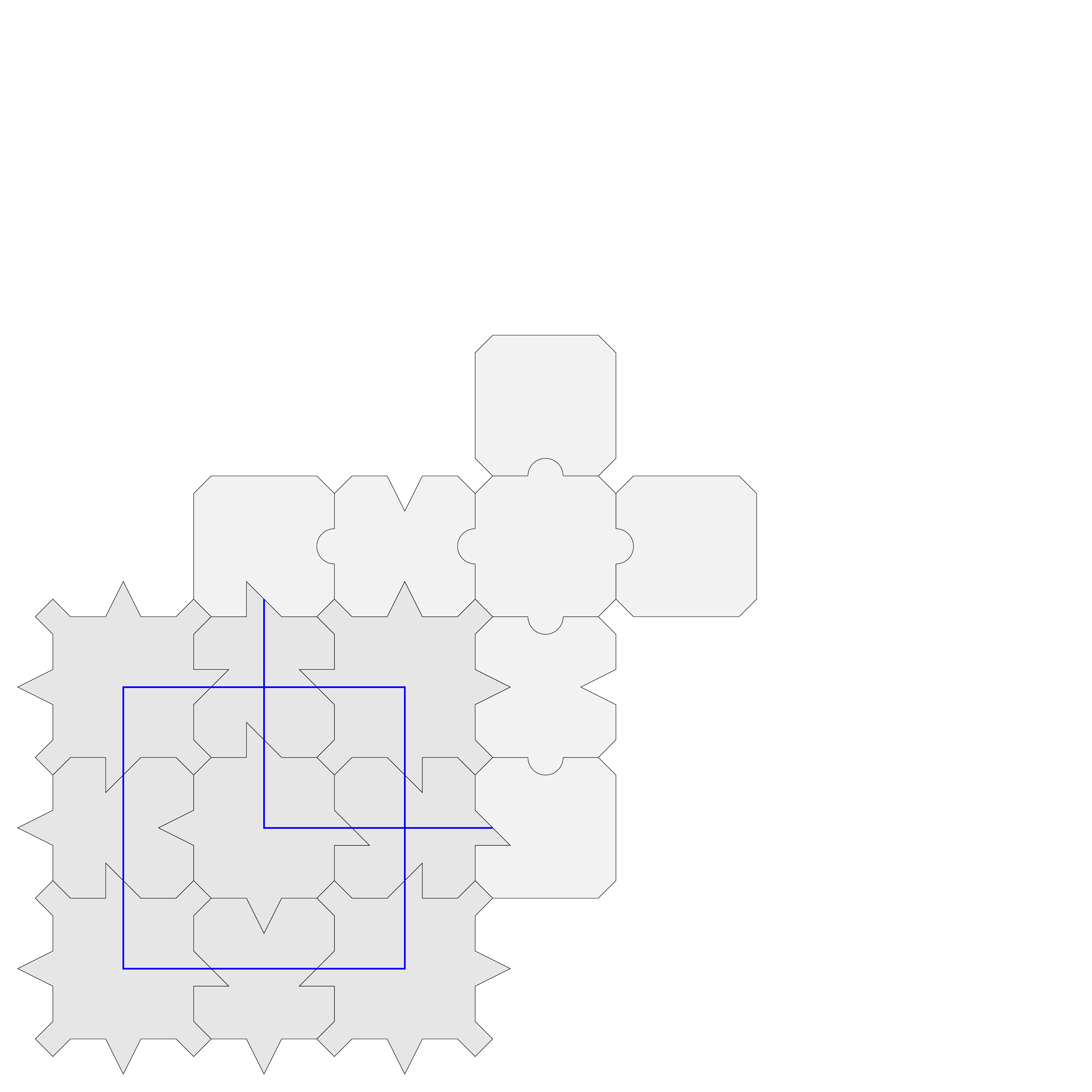}}\hfill
\subfigure[and three order $1$ bumpy corners which appear in order $n$ ones (induction).]{\includegraphics[width=0.2\textwidth]{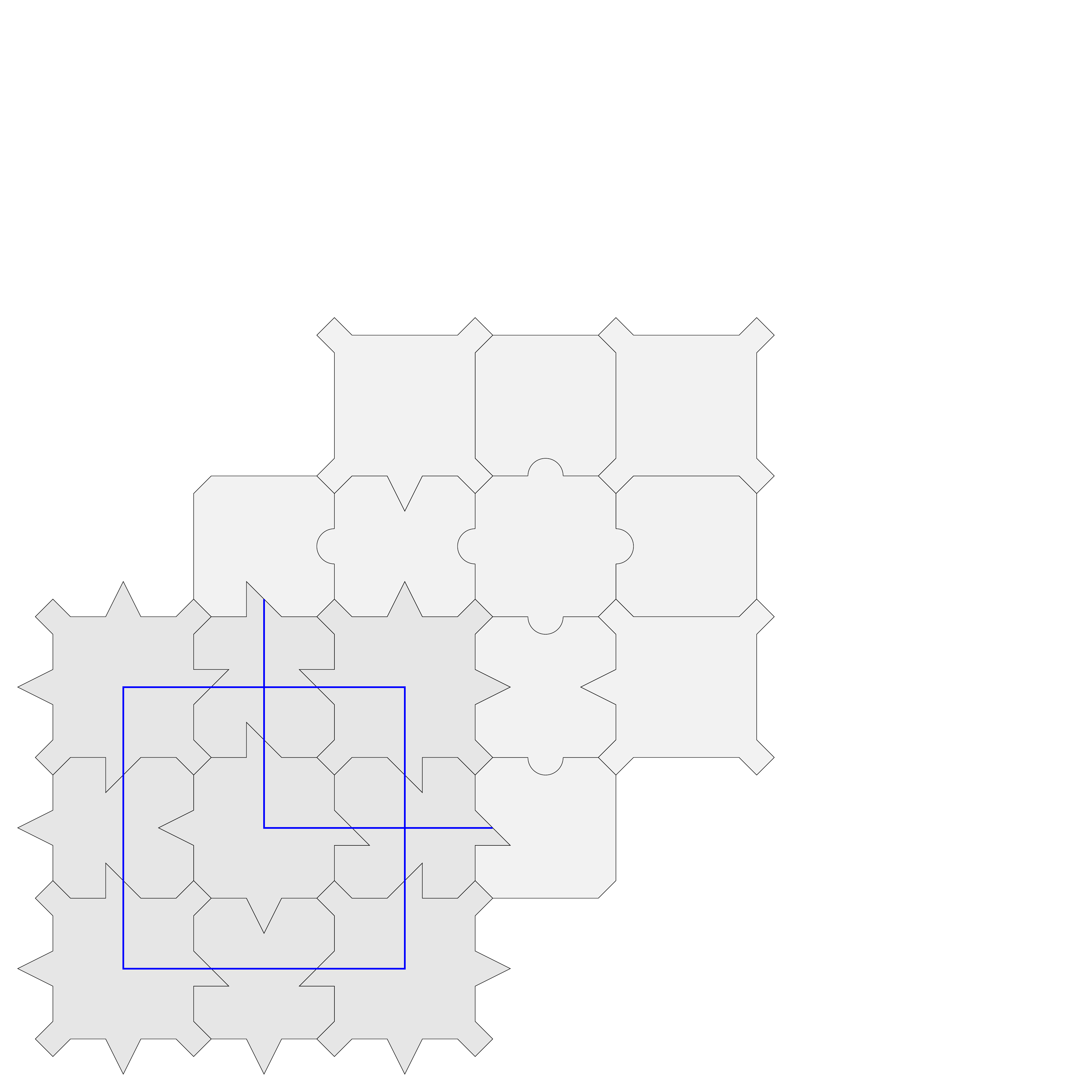}}\\
\subfigure[Their position is forced by their outwards arrows.]{\includegraphics[width=0.2\textwidth]{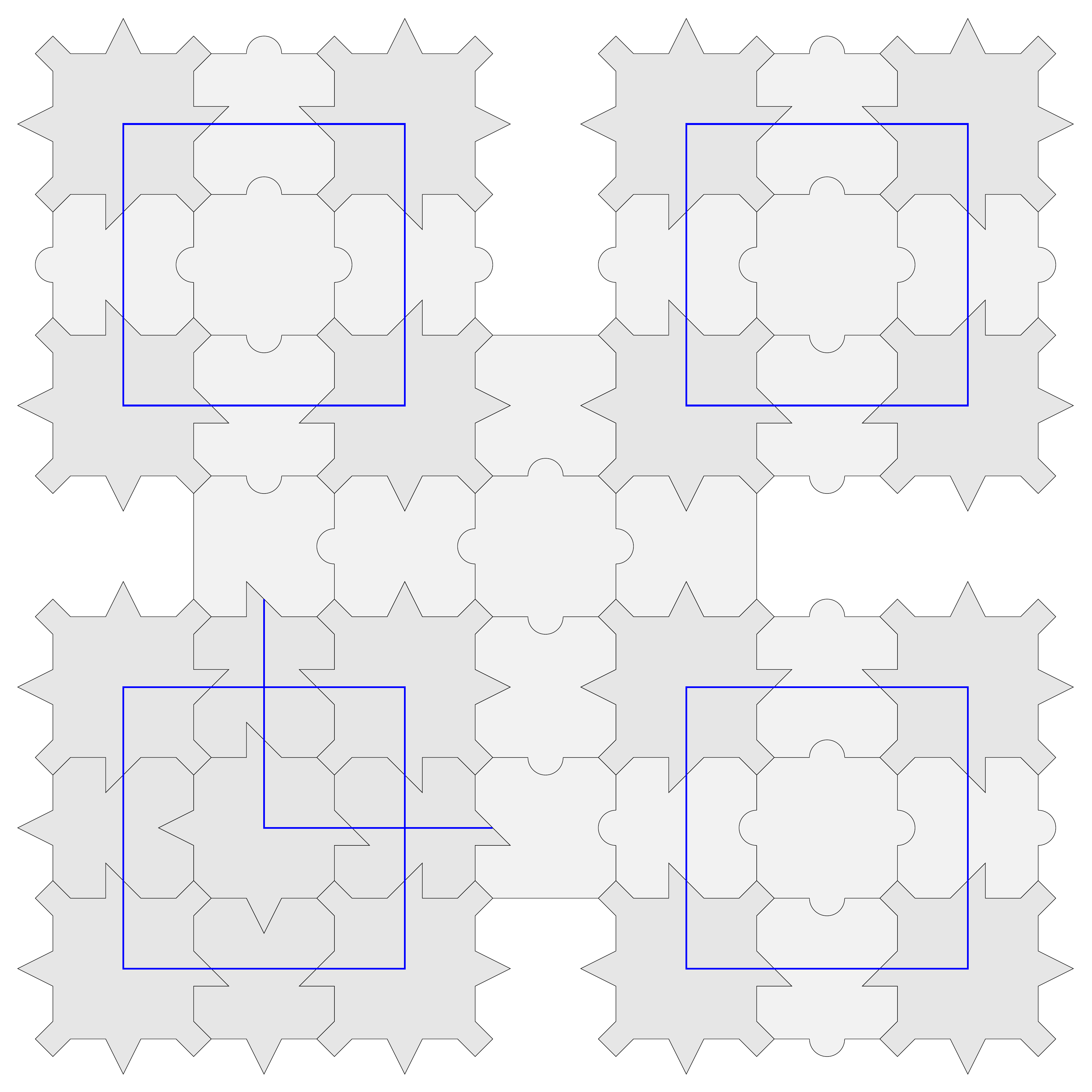}}\hfill
\subfigure[Gaps must then be filled by arms oriented away from the central corner.]{\includegraphics[width=0.2\textwidth]{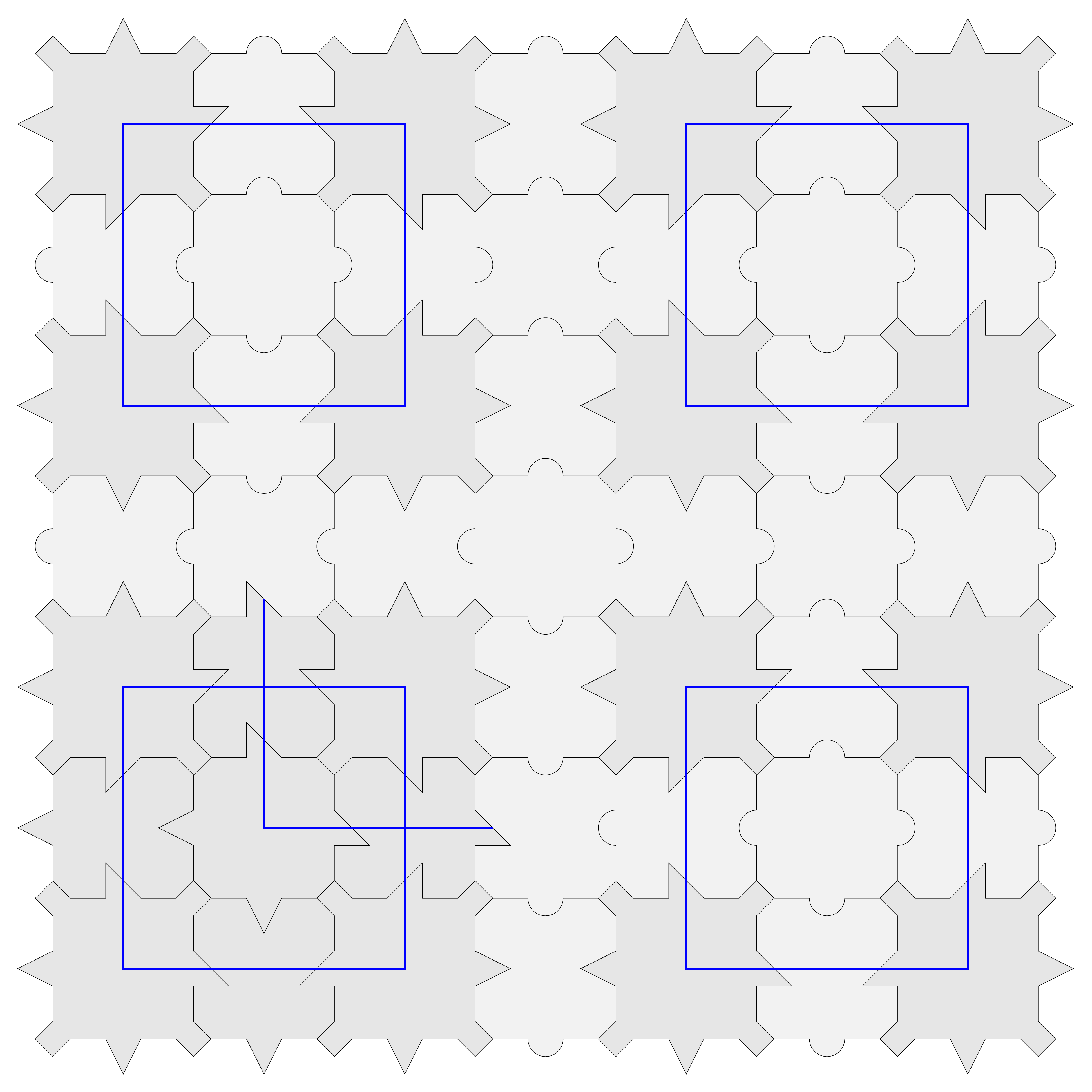}}\hfill
\subfigure[This fixes the orientation of all the order $n$ bumpy corners.]{\includegraphics[width=0.2\textwidth]{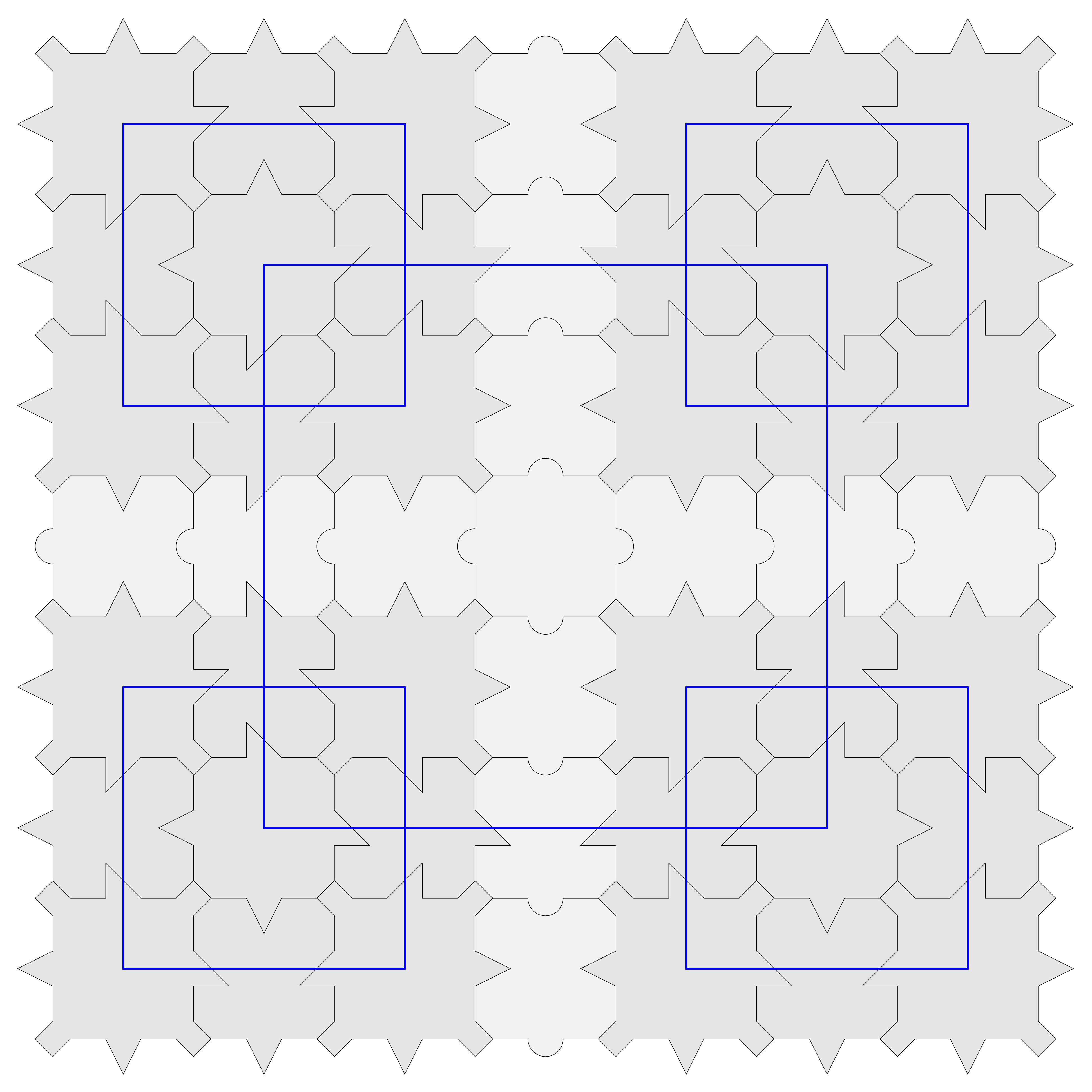}}\hfill
\subfigure[Last, orienting the central corner fixes the arrow types of all the arms.]{\includegraphics[width=0.2\textwidth]{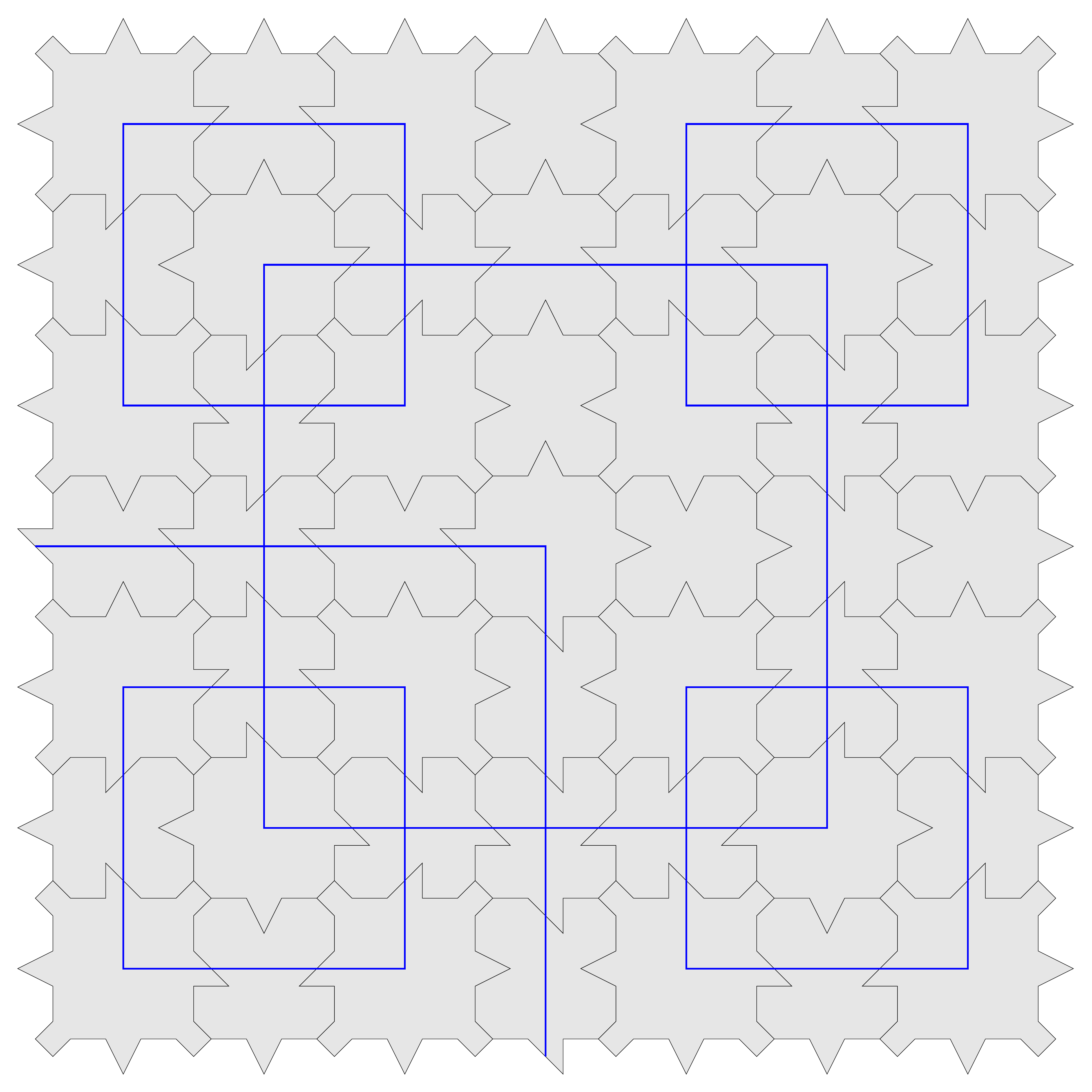}}
\caption{\small Proof that the order $n\geq 1$ bumpy corners of a Robinson tiling appear in order $n+1$ bumpy corners (left-to-right and top-to-bottom, here with $n=2$).
An inwards or outwards arrow whose type is unknown is represented by a dented or bumpy half-disk, accordingly.
An arrow whose orientation is unknown is not represented at all.
Tiles are shaded lighter while their arrows have not all been determined.
}
\label{fig:embedded_bumpy_corners}
\end{figure}

On the other hand, consider a Robinson tiling.
It clearly has order $1$ bumpy corners.
One then inductively proves that \emph{all} its order $n\geq 1$ bumpy corners \emph{necessarily} appear in its order $n+1$ bumpy corners (Fig.~\ref{fig:embedded_bumpy_corners}).
Hence, it has bumpy corners of arbitrary order, on which are drawn arbitrarily large blue squares that no translation can leave all invariant (it should be arbitrarily large!).
Thus, no Robinson tiling is periodic.\\

\noindent The aperiodicity of the Robinson six-tile set follows.

\renewcommand*{\thesubfigure}{}
\setlength{\subfiglabelskip}{0pt}

\end{document}